\documentstyle[titlepage,12pt,psfig]{article}

\def\be{\begin{equation}}

\def\ee{\end{equation}}

\newcount\sectionnumber
\def\sect
{\global\equationnumber=0
\global\advance\sectionnumber by 1
\the\sectionnumber . }
\newcount\equationnumber
\def   \num
{\eqno{\global\advance\equationnumber by 1
\left(\the\sectionnumber .\the\equationnumber \right)}}
\begin{document}

\title{Stable central structures in topologically nontrivial Anti-de
Sitter spacetimes} 

\author{W. L. Smith and R. B. Mann\\
        Department of Physics\\
        University of Waterloo\\
        Waterloo, Ontario\\
        N2L 3G1}

\date{June 2, 1998\\}
\maketitle

\begin{abstract}

We investigate stable central structures in multiply-connected, anti de
Sitter spacetimes with spherical, planar and hyperbolic geometries.
We obtain an exact solution for the pressure in terms of the radius
when the density is constant.
We find that, apart from the usual simply-connected spherically symmetric
star with a well-behaved metric at $r=0$, the only solutions with non-singular
pressure and density have a wormhole topology. However these wormhole solutions
must be composed of matter which violates the weak energy condition. Admitting
this type of matter, we obtain a structure which is
maintained via a balance between its cohesive tension and its repulsive negative
matter density.  If the tension is insufficiently large, this structure can
collapse to a black hole of negative mass.
\end{abstract}

\section{Introduction}

Multiply-connected spacetimes are attracting an increasing amount
of attention among gravitational physicists and cosmologists.
Although the idea that our universe could be topologically non-trivial
has been around for quite some time \cite{topuni}, 
the possibility of setting observational constraints on this
topology by performing a careful search for particular 
correlations in the cosmic microwave background is a recent development
\cite{Weeks}.  Further interest has been spurred
by the realization that domain walls in the early universe can give
rise to pair-production of black holes with event horizons whose topology
is non-trivial \cite{mannlcqg}.  Such objects have been referred to
as Topological Black Holes, or TBHs.

It is the formation of these objects that we are concerned with in this paper.  In
an earlier paper, we demonstrated that a dust cloud in a
multiply-connected, anti de Sitter spacetime could collapse to a TBH in a
process that is analogous to the usual Oppenheimer-Snyder collapse
\cite{us}. The solutions obtained matched a static exterior spacetime to a
dynamic collapsing cloud. The question of whether corresponding static
solutions exist was left unresolved.  

In this paper, we investigate the existence of stable perfect fluid
solutions in multiply-connected, anti de Sitter spacetime.  The spatial
sections of such spacetimes have the topology $R\times H_g$ where
$H_g$ is a two-dimensional compact space of genus $g$.  Such a space
may be described by a metric which is either
flat, spherically or hyperbolically symmetric.  The spherically symmetric
case is simply connected and has genus $g=0$.  The
flat and hyperbolic cases are made compact via appropriate identifications
in those two dimensions.  Perfect fluid solutions in such spacetimes are
centrally located, separated from the exterior spacetime at a constant
radius.  In the $g=0$ case they correspond to a ball of fluid surrounded
by a cosmological vacuum spacetime; in the $g\neq 0$ case the analogous objects
may be referred to as topological stars.  

We find that the only stable solutions for $g > 0$ necessitate the development
of a wormhole inside the star. A constant positive matter density 
throughout the star necessarily implies an infinite pressure somewhere in its 
interior, excluding such objects as stable solutions.  

The situation is considerably different if we consider matter which
violates the weak energy condition.  It has been shown that negative
concentrations of stress-energy can collapse to black holes of negative
mass provided the cosmological constant is sufficiently large in magnitude
\cite{negmass}.  We find that it is possible to construct topological
(wormhole) stars with constant negative matter density 
and have finite negative pressure throughout. The negative pressure
is a tension which acts to hold together the negative density, which is
gravitationally self-repulsive.

Section 2 discusses the three spacetimes of interest, as well as their
topologies.  The spherically symmetric case has much in common with earlier
studies, but we dropped the demand that the metric be well-behaved at the
origin.  Section 3 introduces the interior metrics and their Einstein
equations.  In section 4, the structure of the solutions for a variable
matter density is found using perturbative techniques.  Section 5 produces
a solution for the pressure within wormholes of constant density as a
function of radius.  It is shown in section 6 that this must necessarily
become infinite somewhere within a wormhole with constant positive density. 
The question of wormholes with negative matter density is addressed in
section 7. 

\section{Fluid Topology}

The universe described by the static solution presented here is an
asymptotically anti de Sitter spacetime with a nontrivial topology.   
The exterior metric in this universe, adapted from \cite{bh7}
\begin{eqnarray}\label{extmet}
ds^2 & = & -\left( - {\Lambda \over 3G}\hat R^2 +b - {2M \over \hat R}
\right) dT^2 +  {d \hat R^2 \over - {\Lambda \over 3}\hat R^2 +b - {2M
\over \hat R}} + R^2 (d \hat \theta ^2 + s(b, \hat \theta) ^2 d\hat \phi^2)
\cr \cr 
& &s(b, \hat\theta )  =  \cases{ 
 \sin( \hat \theta ),  & if $b$ = +1  \cr
 1,              & if $b$ = 0\  \cr
 \sinh(\hat  \theta ), & if  $b$ = -1  \cr }
\end{eqnarray}
where $T$ and $R$ are the time and radial coordinates.  $M$ represents the mass of
the star and $\Lambda$ is the cosmological constant (where $\Lambda > 0$ corresponds to
the de Sitter case). $\hat{\phi}$ assumes values between $0$ and $2 \pi$.  

When $ b = + 1$, the universe takes on the familiar spherically symmetric
form, and the $(\hat{\theta}, \hat{\phi})$ sector has constant positive
curvature.  The cosmological constant may assume either sign, but we will
only examine behavior resulting from negative values in the models explored
here.  When $b=0$, the space is flat, with $\Lambda$ less than zero.  In
order to produce the central star to be examined, a space with two sides
identified is considered.  The outer edge of the star will be an identified
flat plane of constant `radius,' $R$. Identification requires the edges of
this plane to be geodesics, which are in this case straight lines.  The sum
of the angles must be $2 \pi$, yielding a parallelogram, or in the simplest case, 
a square or rectangle with opposite
sides identified, creating a toroidal topology, as in figure 1A.  The
exterior universe will maintain this topology with two compact and one
infinite spatial dimensions.

When $b=-1$, the $(\hat{\theta}, \hat{\phi})$ sector is a space with
constant negative curvature, also known as a hyperbolic plane, a
discussion of which is available in Balasz and Voros \cite{bala}.
Geodesics are intersections between the hyperbolic plane and planes
through the origin. A compact surface is formed from the hyperbolic plane
by identifying opposite sides of a suitable polygon whose edges are
geodesics.  The polygon must have a minimum of eight sides, and the
number of sides must be a multiple of four to avoid conical singularities.
An identified polygon with
$4g$ sides is of genus $g$.  The genus determines the topology of the
compact space.  The surface genus $g=2$ has two `holes', and so is a
double-holed doughnut or a pacifier.  The surface with $g=3$ has three
`holes' and has a pretzel shape, and so on, as shown in figure 1. This
type of identification of hyperbolic surfaces is described in more detail
in \cite{us}.

\begin{figure} \label{F1}
\centerline{\psfig{figure=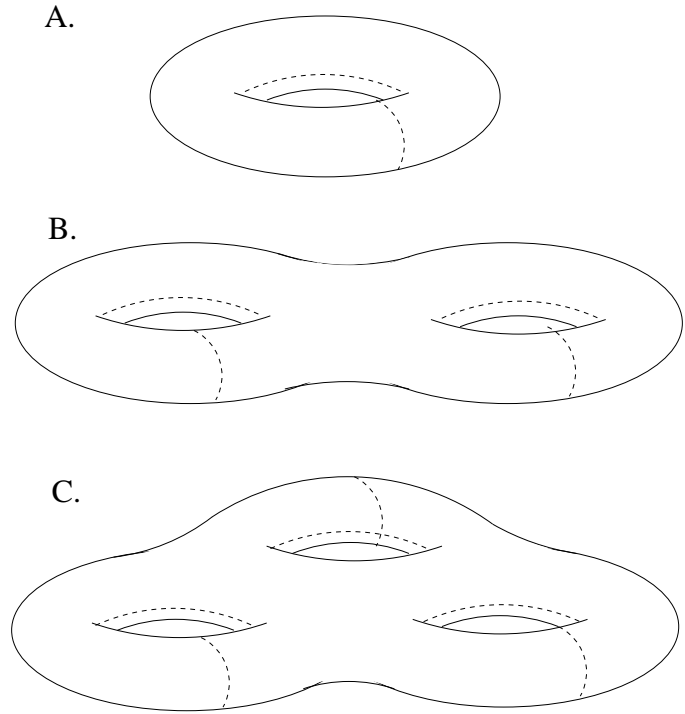,height=8cm,angle=270}}
\caption{Examples of higher genus two-surfaces.  The dotted lines
represent the identifications of edges. A. The torus, $g=1$, is a
representation of a square or rectangle in flat space identified in two
dimensions. B. The pseudosphere, $g=2$, is the simplest appropriate
hyperbolic two-surface, an octagon, after identification.  C. This $g=3$
surface is a slightly more complicated possibility.} \end{figure}

The fluid cloud in these identified spacetimes is located in a central
position, analogous to the central sphere in a spherically symmetric
spacetime.  The boundary between the fluid and the exterior universe is an
identified flat or hyperbolic plane.  The universe outside the fluid
maintains the same topology of the cloud boundary itself.  The fluid forms
a pacifier within a pacifier with the holes lined up, or the banana cream
in our doughnut, if you will.  The situation is analogous for higher genus
topologies.  Beyond the cloud, the radial coordinate may range to infinity.

\section{Calculations of the Einstein Equations}

	The standard metric of an arbitrary, static spacetime with
spherical, toroidal or hyperbolic symmetry was used in conjunction with the
Einstein equations to generate interior solutions.  The metric is given by: 
\begin{eqnarray}
ds^2 & = & -F(r,b) dt^2  + H(r, b) dr^2 + r^2 (d \theta ^2 + s(b, \theta)
^2 d\phi^2),
\end{eqnarray}
with $s(b, \theta )$ as given in (\ref{extmet}) above.
Here, $t$ is the time coordinate, $r$ is the radial coordinate, and
$\theta$ and $\phi$ are coordinates on a two-surface of constant positive,
zero or negative curvature, where $\phi$ has a range of $0$ to $2 \pi$.
The matching of the metrics is analogous to that in \cite{us}, and is
carried out under the conditions that the metrics and the extrinsic
curvatures match smoothly across the boundary.

The cosmological constant will be absorbed into definitions of the density
and pressure, so that
\be \rho = \rho _m - { | \Lambda | \over 8 \pi G}, \qquad 
P = P_m + {| \Lambda | \over 8 \pi G},
\ee
where $\rho_m$ and $P_m$ are the density and pressure due to matter
respectively.  Since the $\Lambda < 0$ case is of interest here, the
absolute value of $\Lambda$ is used for the remainder of this paper. 
Einstein equations are therefore simply
\be
G_{\mu \nu} = - 8 \pi G T_{\mu \nu}
\ee
\be
 T_{\mu \nu} =P g_{\mu \nu} + ( \rho + P)u_\mu u_\nu 
\ee
with the fluid 4-velocity given by $u^\mu = - \sqrt{F} (dt)^\mu$.

This means that the components of the Einstein tensor are,
\be \label{Gtt}
G_{00}=-F \left( {H^\prime \over H^2 r} -{1 \over r^2} \left(1/H -b \right) 
\right)=-8 \pi \rho F
\ee
\be \label{Grr}
G_{11}=-H \left( {F^\prime \over F H r} +{1 \over r^2} \left(1/H -b \right) 
\right)=-8 \pi P H 
\ee

\begin{eqnarray} \label{Gthth}
G_{22}  &  = & -{r \over 4 F^2 H^2} \left(2 F^\prime F H -2 H^\prime F^2+ 
2 r F^{\prime \prime}  F H - r F^{\prime 2} H - r F^\prime H^\prime F\right) 
\cr & = & -8 \pi P r^2 = {G_{33}\over \sinh^2 \theta}
\end{eqnarray}
where the primed variables refer to the derivative, $d/dr$.

\section{Variable density solutions}\label{Vrho}

Equation (\ref{Gtt}) produces the following solution for $H(r)$:
\be\label{Heq1}
H = {r \over  { | \Lambda| \over 3G} r^3 + b r + c - \int^r_{r_m} 8 \pi \rho_m r^2}
\ee
where $c$ is a constant of undetermined sign, and $r_m$ is the minimum value
of the radius.  In order to examine behavior of the metric, expand about the
minimum radius with the infinitesimal, $r= r_m + \epsilon$, and use a Taylor
series approximation for the integral.  The spatial metric then becomes
\be
ds^2_s = {(r_m + \epsilon ) d \epsilon^2 \over A + B \epsilon
+C \epsilon^2 + D \epsilon^3 + ...} + (r_m+\epsilon )^2 d \Omega^2.
\ee
where $d \Omega^2$ is the appropriate angular spatial section.
The behavior at small $r$ will depend on the parameters, $A$, $B$,
$C$ and $D$, as well as the value of $r_m$.  The parameters are given
by
\begin{eqnarray}
&A  = {|\Lambda |\over 3 G }r_m^3 +b r_m + c, 
& C   = {\textstyle{|\Lambda |\over G}} r_m -8 \pi \rho_m r_m - 4 \pi \rho_m^\prime
r_m^2,
\\
&B  = {|\Lambda | \over G} r_m^2 +b -8 \pi \rho_m r_m^2,
& D  = {\textstyle{|\Lambda | \over 3 G }- {8 \over 3} \pi \rho_m- {16 \over 3}\pi 
\rho_m^\prime r_m - {4 \over 3} \pi \rho_m^{\prime \prime} r_m^2}.\nonumber
\end{eqnarray}

Small $r$ behavior is tabulated in Table 1.  A finite throat refers to the
situation in which the radial coordinate reaches a minimal value at
some finite distance from the outer edge of the fluid,
and then expands again into another universe. In the genus $g=0$ case 
a series of spheres of smaller and smaller proper radii are encountered, as an
observer travels into the star.  Eventually a sphere of minimum size is
encountered, beyond which the spheres begin to grow once more. In
the genus $g>0$ case, the situation is the same, except that the
spheres are replaced with pseudospheres (or planes) which are identified under
the action of a discrete isometry group, as described in section 2.  Hence
the spacetime within the fluid has a wormhole structure, and may be
matched at each end of the wormhole (where $r=R$) to an exterior spacetime
whose metric is given by (\ref{extmet}).

In the infinite trumpet, the wormhole is infinitely long; \i.e. the center
of the star is an infinite proper distance from its surface. The infinite
hourglass is a wormhole of infinite proper length connecting two spaces
where the radius may grow to infinity, \i.e. two universes. 

From table 1, it is readily apparent that the interesting cases, the
finite throats, are those for which the behavior of the metric at small
$r$ is independent of $\rho^\prime$.  We
will next examine the constant density case in more detail to see if the
solutions are indeed stable.

\begin{table}[t]{
\begin{tabular}
{ |c||c|c|c|c| }\hline 
 & $A \ne 0$
& $A = 0 $, $B \ne 0 $ 
& $A = B = 0 $, $C \ne 0 $ 
& $A = B = C = 0 $, $D \ne 0 $ 
\\  \hline \hline
 $r_m = 0$ & cusp & no solution$^\romannumeral1 $ & no solution &
infinite hourglass$^{\romannumeral3}$
\\ \hline
Ricci scalar& $2b ({3\over 2} c \ell)^{-4/3}$&N/A$^{\romannumeral2}$
&N/A&$-6D$
\\ \hline 
 $ r_m \ne 0 $
& finite throat & finite throat & infinite trumpet &
infinite trumpet
\\ \hline
Ricci scalar& $2b/r_m^2$ & $2b/r_m^2$ & $2b/r_m^2$ & $2b/r_m^2$  
\\  \hline 
\end{tabular} }    

\caption{ A summary of small $r$ behavior for stars with variable $\rho
_m$.  When $r_m $ is non-zero, the Ricci scalar becomes $2b/r_m^2$, while
in the cusp case, it is inversely proportional to the proper length, 
$\ell$, to the four-thirds power.  For $A=0$ and $B\neq 0$ (second column)
when $r_m=0$ a
spherically symmetric star in an anti de Sitter spacetime may form,
 with the Ricci scalar vanishing at the center. 
For the infinite hourglass (fourth column) only $b=0$ is allowed.  In
every case, the relevant nonzero constant $A, B, C, or D$ must be
positive, restricting the possible values of $\rho_m$ and $|\Lambda |$.}

\end{table}

\section{Constant Density Stars and the Buchdahl Identity}

Equation (\ref{Gtt}) leads to a solution for the function $H$, if the
density is taken as constant,
\be \label{Heq}
H= \Biggl( \beta r^2 +b+ { \alpha \over r} \Biggr) ^{-1}
\ee
in which $\alpha$ an arbitrary constant and
\be 
\beta= -{8 \over 3} \pi \rho =-{8 \over 3} \pi \rho_m + {|\Lambda | \over 3G}.
\ee
Here, $\rho$ is the net density.  For now, we assume the matter density is
positive or zero.  $H$ must always be greater than zero, to preserve the
signature of the metric.

If we let $ F = e^{2 \Phi}$, then equation (\ref{Grr}) will be
\be
{d \Phi \over dr} = {8 \pi P r^3 -\beta r^3 - \alpha \over 
2 r (\beta r^3 + b r+ \alpha )}
\ee
The final Einstein tensor equation, (\ref{Gthth}), along with equation
(\ref{Heq}) leads to a solution for the change in pressure as a function of
$r$: 
\be
{dP \over dr}=-(P+\rho) {d \Phi \over dr} 
= -{(8 \pi P - 3\beta )(8 \pi P r^3 -\beta r^3 - \alpha) \over 
16 \pi r ( \beta r^3+ br+\alpha )}
\ee

In order to solve this equation, note that it can be put in the form
\be 
\Biggl[ {16 \pi \over r} {\left( \beta r^3 +b + \alpha \right)\over 8 \pi P
-3 \beta}  \Biggr] dP + \Biggl[ { 1 \over r^2} 
\left(8 \pi P r^3 - \beta r^3 - \alpha \right)\Biggr] dr = 0.
\ee
and the integrating factor
\be
\mu = \left({r \over \beta r^3+br + \alpha} \right) ^{3/2} { -1 \over 8 \pi
P - 3 \beta}.  \ee
applied.  By integrating from the outer edge of the star, where the pressure is
$| \Lambda | /8 \pi G$ to an arbitrary radius within the star, we
find that
\be \label{Psol}
P = {\rho_m  \over \sqrt{(\beta r^2 +b +\alpha /r)\over ( \beta R^2 +b +
\alpha /R)} - 4 \pi \rho_m \sqrt{(\beta r^2 +b +\alpha /r)} \displaystyle
\int  _r^R {r dr \over  ( \beta \tilde{r}^2 +b + \alpha / \tilde {r}) ^{3
\over 2}} } +  {3 \beta \over 8 \pi}.
\ee
The condition that $H$ be real for all radii, forces the pressure to be
everywhere real.

The Buchdahl identity is normally found by demanding the central pressure
be finite in a simply connected spacetime, where $g=0$. Furthermore 
$\alpha $ is assumed to vanish, allowing the metric to be well-behaved 
at $r=0$.  In this case it is straightforward to explicitly carry out
the integral in (\ref{Psol}) to obtain
\begin{eqnarray} 
&P ={\displaystyle \rho \Biggl[{K \sqrt{1-{8 \over 3} \pi \rho R^2} -
 \sqrt{1-{8 \over 3} \pi \rho r^2}
\over
\sqrt{1-{8 \over 3} \pi \rho r^2} - 3K \sqrt{1-{8 \over 3} \pi \rho R^2}}
\Biggr],} \cr
&K ={\displaystyle { P(R) + \rho \over 3 P(R) + \rho} }
={\displaystyle { 4 \pi \rho_m \over 4 \pi \rho_m +| \Lambda | / G} \qquad .}
\end{eqnarray}
The pressure should be positive definite, meaning
\be
{1\over 3} \le K \sqrt{1-{8 \over 3} \pi \rho R^2 \over 1-{8 \over 3} \pi
\rho r^2} < 1
\ee
The right hand inequality constrains $K$ most strongly at the edge of the
cloud, and the left at the center.  This second constraint may be
translated as a limit on the mass in terms of the radius, 
\be
M = \int_0^R 4 \pi \rho r^2 dr > {9 K^2 - 1 \over 18 K^2} R
\ee
which reduces to the familiar $M> 4 R/9$ limit when $\Lambda = 0$
\cite{buch}.  By demanding that ${ dP_m\over d\rho_m} \ge 0$, that $M(0) =
0$ and that the pressure was non-negative and bounded everywhere, Hiscock
was able to obtain the stronger constraint
\be
M/R \le {2 \over 9} \left[1- {3 | \Lambda | \over 4G} R^2 + (1+ {3 | \Lambda |
\over 4G} R^2 )^{1/2} \right]
\ee
for the $b=+1$ case. \cite{hiscock}

We will not demand $\alpha = 0$.  The $b=+1$ case has already been
examined for nonzero $\alpha$, in which stable stars with $g=0$ may form,
by Hiscock.  Consider now the higher-genus cases.  When $g=0$, $b=0$ and
the parameter $\beta$ is forced to be positive. When $g\geq 2$, $b=-1$,
implying $\beta > 1/ R^2$.  The equation for the pressure becomes, for
any genus,
\be P = \rho \left( { \sqrt{ \beta r^2 +b } - K
\sqrt { \beta R^2 + b } \over 3 K \sqrt { \beta R^2 + b } - \sqrt{ \beta
r^2 + b } } \right), \quad 
\ee 
provided $\alpha=0$. The analogous Buchdahl identity, found by demanding that
the pressure is positive definite, is then 
\be 
1/3 < K \sqrt{ \beta R^2+b \over \beta r^2 +b} \le 1 
\ee 
for all $r$.  

When $R=r$, the left hand equality demands that $| \Lambda |
/G < 8 \pi \rho_m$ but maintaining a metric with the correct signature
throughout requires $|\Lambda | /G > 8 \pi \rho_m$.  This contradiction
rules out the possibility of a genus $g > 0$ stable star with $\alpha=0$ and
positive pressure everywhere.  

\section{Constant density solutions}

When $\rho$ is constant, the parameters $A$, $B$, $C$ and $D$ from section
\ref{Vrho} may still be nonzero.  The behavior of the solution will depend
on the character of the lower cutoff of the positive region of $\beta r^3 +
b r + \alpha$ being examined.  If that function is cut off by a double or
triple root, an infinite throat will result.  If the region in which the
function if positive is ended by the y-axis, a cusp will be formed.  If,
however, the function has a non-degenerate, positive root as the lower
bound to its positive region, the star will have a finite throat, with a
minimum radius given by that root. 

None of these wormholes will have a well behaved pressure for a positive matter
density.  Note that we can rewrite (\ref{Psol}) as
\be \label{Psol2}
P_m = {\rho_m  \over \sqrt{(\beta r^2 +b +\alpha /r)\over ( \beta R^2 +b + \alpha /R)}
- 4 \pi \rho_m \sqrt{(\beta r^2 +b +\alpha /r)} \displaystyle \int  _r^R {\tilde{r} 
d\tilde{r} \over ( \beta \tilde{r}^2 +b + \alpha / \tilde {r}) ^{3 \over 2}} } - \rho_m.
\ee
The first term in the denominator varies from a value of unity at the
outer radius to zero at the throat.  The integral in the second term is
a well-behaved positive function, so the second term will be zero at the
outer radius and some positive number at the throat.  To see this,
approximate the behavior of the term near a single root, $r_0$.
\begin{eqnarray}
\sqrt{\beta r^2 + b + \alpha /r} \displaystyle \int  _r^R 
{r dr \over (\beta r^2 + b + \alpha /r)^{3/2}}
&\approx& {r_0^2 \sqrt{r-r_0} \over qr_0^2+sr_0+t} 
\displaystyle \int  _r^R {dr \over (r-r_0)^{3/2}} \cr
&\approx&{2r_0^2 \over (qr_0^2+sr_0+t)}.
\end{eqnarray}

It is therefore unavoidable that the two terms in the denominator
will become equal at some value of $r$, at which point the
pressure will be infinite.  This type of wormhole can never be stable.
Sample plots displaying this behavior were obtained by numerical
integration, and are shown in figure 2.

 \begin{figure} \label{F1a}
 \centerline{\psfig{figure=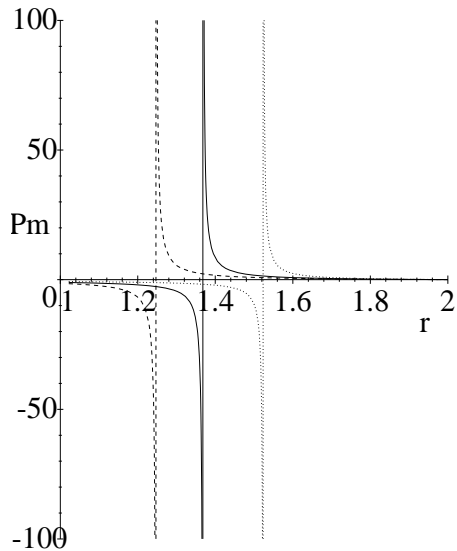,height=4cm,angle=0}} 
 \caption{The pressure, as found in (\ref{Psol2}), is evaluated for the
 special case where the largest root of $H(r,b)$ is a single root, $r_m =1$.
   The equation used is $
  H(r,b)= -{1 \over x} ({8 \over 3} \pi \rho_m - {|\Lambda | \over 3})r_m^2 
 (x-1)  [x^2+x+1 - b [r_m^2 ({8 \over 3} \pi \rho_m- {|\Lambda | \over 3})]
 ^{-1}]$. All the curves have $|\Lambda | = 1$, $\rho_m=.5$, $R=2$ and
 $r_m=1$.  The dashed curve has $b=-1$, the solid curve has $b= 0$, and the
 dotted curve has $b=+1$}
 \end{figure}

\section{Negative matter density wormholes}\label{negrhom}

Wormhole solutions have long been known to require the existence of exotic
matter (matter which violates the weak energy condition)
\cite{Mike}, so it is not surprising that this case also 
requires such.  A study of wormhole solutions in topologically trivial
spacetimes with nonzero cosmological constant also led to this conclusion,
although the actual conditions on the exotic matter
are modified \cite{delg}. The necessity of exotic matter
is an unpleasant but not prohibitive situation, the Casimir effect being
perhaps the best known example of a manifestation of the violation of
the energy conditions.  

The most likely situation in which 
topological black holes have physical relevance is in the early universe 
\cite{bh7} is also one in which quantum fluctuations 
may produce (temporarily at least) regions in which the weak energy condition is
violated. It is therefore natural to consider in more detail topological 
``stars'' in which the energy conditions are violated. 
Indeed, a study of a dust cloud of negative energy density
indicated that exotic matter may behave in a counter-intuitive
manner, collapsing to form black holes \cite{negmass}.

The simplest case which requires $\rho_m$ to be negative is that for which
the parameter $\alpha $ vanishes, as referred to earlier.  The behavior
of the pressure in this situation is representative of the more complicated 
$\alpha \neq 0$ cases. Here the matter density is forced to be negative
in order that the metric be real.  The pressure takes on the simple form
\be
P_m =- |\rho_m | \left({ \sqrt{\beta r^2 +b} - \sqrt{\beta R^2 +b} \over {3 | \rho_m | \
\over | \rho_m | - | \Lambda| / ( 4 \pi G) } \sqrt{\beta R^2 +b} -\sqrt{\beta r^2 +b}}
  \right).
\ee
In this case the pressure is well-behaved. It vanishes at the star's
edge and decreases to a finite central value at $r = r_m$.  The pressure
here is always negative, so in fact it is a tension.  
Generalizing the constraint from ref.\cite{hiscock} 
to include negative matter pressure and density yields
\be\label{hcon}
{ dP_m \over d \rho_m} \ge 0 \ge {dP_m \over d |\rho_m |}.
\ee
This will guarantee that the matter pressure $P_m$ must become more 
negative as the matter density becomes more negative, as is a physically
reasonable.  This constraint may also be interpreted as a
limit on the matter density, since when it becomes too negative the
inequality will no longer hold throughout the star.  This constraint is
most easily evaluated numerically.

\begin{figure}\label{F3a}
\centerline{\psfig{figure=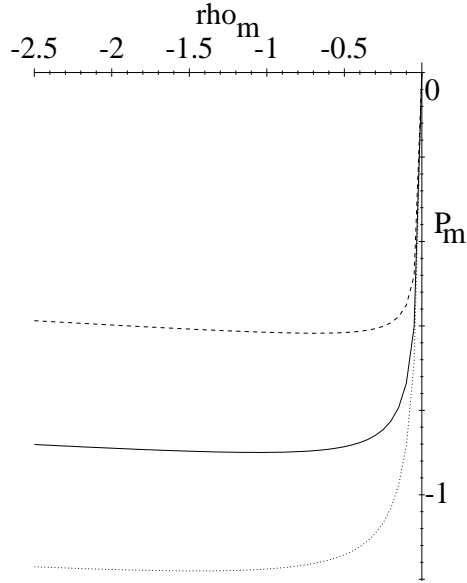,height=4cm,angle=0}} \caption{The matter
pressure as a function of matter density for $b=0, +1, -1$, represented by
the solid, dotted and dashed lines.  All plots use $|\Lambda | /G = 1$, $r
=1.001$, $R=2$ and $r_m=1$ in the equation show for figure 2.  The cutoff of
the allowed region is at $\rho_m = -1.06, -1.45,$ and $-.69$ for $b=0, +1$
and $-1$ respectively.} \end{figure}

The behavior of the matter pressure when $\alpha$ is non-zero is
qualitatively similar.  The function is well-behaved, zero at the outer edge
and finite at the throat.  The same constraint (\ref{hcon}) as before is
employed to produce physically reasonable results.  The strongest constraint
arises at a radius near the throat radius, as was done in figure 3.

\begin{figure} \label{F3} 
\centerline{\psfig{figure=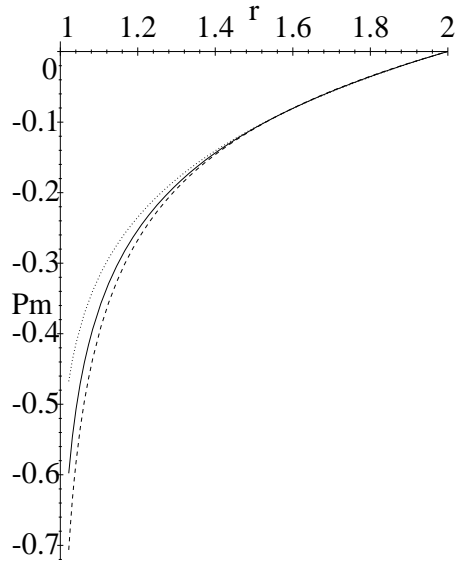,height=4cm,angle=0}} 
\caption{The matter pressure, as found in (\ref{Psol2}), is evaluated for the
same equation for $H(r,b)$ as before, but now $\rho_m<0$ is permitted.  The
parameters used to find the solid, dotted and dashed lines are $b=0, +1,
-1$.  All plots use $ |\Lambda|/G = 1$, $\rho_m=-.5$, $R=2$ and $r_m=1$.
The lines here approach finite values at the throat, $r=r_m$.}

\end{figure}

\section{Conclusions}

Although the collapse of a pressureless dust cloud of positive energy to a topological
black hole proceeds in a manner somewhat analogous that in the usual spherical
case (with genus $g=0$) \cite{us}, the formation of stable central structures in
topologically non-trivial anti de Sitter spacetimes differs considerably from the
topologically trivial case. Indeed geometric requirements are in conflict with
energy positivity requirements, implying that there are no stable central structures
formed from a perfect fluid respecting the energy conditions and
whose exterior metric is given by (\ref{extmet}).   

The only ``stable'' solutions are those in which 
the matter density within the star is negative, with a magnitude smaller than a 
critical value determined by (\ref{hcon}).  In the case the topological star consists
of a fluid of gravitationally repulsive negative energy, held together by a sufficiently
large tension (\i.e. negative pressure). Both pressure and density are finite everywhere
throughout the star.  This is the reverse of a normal star, whose gravitationally
self-attractive density is prevented from collapsing by its pressure.  Should the
pressure of the negative-mass star decrease below a certain threshold during its
evolution, it will either explode due to gravitational self-repulsion
or collapse to a black hole of negative mass.
The former situation will occur if the magnitude of the density is sufficiently large
relative to $|\Lambda|/G$. Otherwise the evolution of the star should
proceed along the lines described in ref. \cite{negmass}, ultimately 
reaching a black hole of negative mass as its final state.

The most likely physical situation in which any of these scenarios is relevant
is in the early universe.  Topological black holes can be formed via pair-production
in the presence of domain walls \cite{bh7} or from the collapse of a dust cloud 
(of either positive or negative density) in a topologically suitable 
setting \cite{us,negmass}. However, if it is possible to produce exotic matter in
such settings, the results of this paper indicate that stable (wormhole-type)
central structures can form.

\section{Acknowledgements}
This work was supported by the National Science and Engineering Council of
Canada.

\end{document}